# Photo-Nernst detection of cyclotron resonance in partially irradiated graphene

Kei Kinoshita[1], Rai Moriya[1,*], Satoru Masubuchi[1], Kenji Watanabe[2], Takashi Taniguchi[2], and Tomoki Machida[1,*]

[1] *Institute of Industrial Science, University of Tokyo, 4-6-1 Komaba, Meguro, Tokyo 153-8505, Japan*

[2] *National Institute for Materials Science, 1-1 Namiki, Tsukuba 305-0044, Japan*

Cyclotron resonance of a Landau-quantized graphene can absorb significant amount of infrared light. However, application of this phenomenon to the photodetector had been limited due to the lack of efficient photon to charge conversion scheme. Here, we demonstrate the detection of cyclotron resonance in a partially metal-masked monolayer graphene two-terminal device using photo-Nernst effect. Due to the presence of the mask, incident infrared light is irradiated on only one-half of the graphene channel. This partial irradiation creates a temperature gradient perpendicular to the graphene channel. In the presence of an external magnetic field, thermopower is generated perpendicular to the temperature gradient due to the Nernst effect. Consequently, photo-Nernst voltage is generated along the graphene channel, which can be detected from the contacts on both ends of the channel. We demonstrate selective detection of the photo-Nernst effect while minimizing the other photovoltaic contributions, such as the photo-Seebeck effect. We investigate the dependence of the photo-Nernst effect on magnetic field and **excitation wavelength, which reveals a significant enhancement of photo-Nernst signal at the cyclotron resonance conditions in graphene. Our finding could facilitate the realization of far-infrared light detector using cyclotron resonance of graphene.**

*E-mail: moriyar@iis.u-tokyo.ac.jp; tmachida@iis.u-tokyo.ac.jp

Graphene has shown immense potential for novel applications in photodetection in the mid-infrared to longer wavelength regime, where a suitable photoconductive material has been lacking [1-3]. Various types of graphene-based infrared, terahertz, and microwave photodetectors have been experimentally investigated [4-9]. Photo-thermoelectric effect has been recognized as the most dominant mechanism of graphene-based photodetectors. [1,10-13]. In this effect, the light absorption in graphene significantly increases its electron temperature owing to the small electron heat capacity of graphene and as the thermoelectric coefficient of graphene is moderately large, this electron heating can be efficiently converted into photo voltage. Previously, we had demonstrated that cyclotron resonance of graphene can be detected by using the photo-thermoelectric effect, and the main contribution of photo-thermoelectric effect is photo-Seebeck effect in two-terminal graphene devices [14]. However, in principle, the photo-thermoelectric effect has another form, called the photo-Nernst effect [15], which has not been studied extensively for infrared photodetection as well as for detection of cyclotron resonance.

The difference between photo-Nernst and photo-Seebeck effect is schematically illustrated in Fig. 1. Figure 1(a) shows the schematic of a graphene two-terminal device exhibiting photo-Nernst effect, where only one half of graphene (bottom half in the figure) is irradiated by using a metal mask. This creates a temperature gradient within the graphene channel such that the bottom side of the channel is hot while other regions of the device are cold (Fig. 1(b)). This implies that the direction of the temperature gradient is perpendicular to the channel. In contrast, the two-terminal device exhibits the photo-Seebeck effect in the absence of the metal mask (Fig. 1(f)). Here, the dominant temperature gradient is between graphene and the metal contact under light irradiation, i.e., graphene is hot and the metal contact is cold, as shown in Fig. 1(g). In this case, the temperature gradient is parallel to the graphene channel.

These different temperature gradient directions are expected to generate different photo-thermoelectric responses under the variation of both carrier density of the graphene $n$ as well as perpendicular magnetic field $B$. Photo-Nernst voltage $V_{NR}$ and photo-Seebeck voltage $V_{SB}$ can be expressed as follows [15,16]:

$$V_{NR} \propto (\sigma^{-1})_{xx}\left(\frac{\partial \sigma}{\partial E_F}\right)_{xy} + (\sigma^{-1})_{xy}\left(\frac{\partial \sigma}{\partial E_F}\right)_{yy}, \quad (1)$$

$$V_{SB} \propto (\sigma^{-1})_{xx}\left(\frac{\partial \sigma}{\partial E_F}\right)_{xx} + (\sigma^{-1})_{xy}\left(\frac{\partial \sigma}{\partial E_F}\right)_{yx}, \quad (2)$$

where $\sigma$ depicts the conductivity tensor and $E_F$ the Fermi energy. The temperature gradient for the Nernst effect is shown in Fig. 1(b), which causes the carriers to move in a direction perpendicular to the channel; therefore, Nernst voltage is not generated at zero magnetic field, as shown in Fig. 1(c). Under the application of the magnetic field, Lorentz force causes lateral displacement of moving carriers; this is known as the Nernst effect. A thermopower is then generated in a direction parallel to the channel, which is detected by the two contacts located at both ends of the graphene. Based on Eq. (1), the dependence of $V_{NR}$ on carrier density $n$ in a large magnetic field is illustrated in Fig. 1(d). $V_{NR}$ exhibits a peak around the Dirac point and approaches zero at large values of $n$. Furthermore, the oscillatory variation of $V_{NR}$ as a function of $n$ is due to the Landau quantization of graphene. Notably, the sign of photo-Nernst voltage changes under the reversal of the magnetic field [15]. The magnetic field dependence of the photo-Nernst voltage at a fixed carrier density is depicted in Fig. 1(e). At low magnetic fields, $V_{NR}$ increases monotonically due to the increase in the Nernst coefficient of graphene [16], while at higher magnetic fields, it starts to oscillate. These are the characteristic features of the photo-Nernst effect.

For the photo-Seebeck effect shown in Figs. 1(f) and 1(g), the carriers move along the temperature gradient and generate thermopower. Thus, the direction of the Seebeck voltage is parallel to the temperature gradient and it is locally generated at the junction between graphene and graphene underneath the metal contact. As the Seebeck voltage is generated in opposite directions at the left- and right-contact regions (Fig. 1(g)), the two components can easily cancel each other. Based on Eq. (2), the photo-Seebeck voltage for single graphene/metal junction is shown in Figs. 1(h-j). Detection of this voltage requires breaking the symmetry between the two contacts. For example, the use of different materials for the right and left metal contacts or local back gate [10,11,14,17] can create different carrier dopings (thus, different Seebeck coefficients) in graphene at the two contacts. $V_{SB}$ is finite even at zero magnetic field and its carrier density dependence is depicted in Fig. 1(h). $V_{SB}$ changes its sign depending on the carrier type, which is the characteristic of the Seebeck effect in graphene. At high magnetic fields (Fig. 1(i)), $V_{SB}$ exhibits oscillatory variation due to the Landau quantization. In contrast to the photo-Nernst effect, $V_{SB}$ does not depend on the sign of the magnetic field. The magnetic field response of $V_{SB}$ is observed to be fully symmetric under field reversal (Fig. 1(j)). We also note from Fig. 1(b) that, in addition to the photo-Nernst effect, the temperature gradient between graphene and the metal contact can cause the photo-Seebeck effect around the metal contact. However, the Seebeck contribution from the two contacts cancel each other and do not influence the Nernst measurement. Therefore, it is evident that the symmetry of induced voltage under the reversal of carrier density and magnetic field is significantly different in the photo-Nernst and photo-Seebeck effects, which can be used to distinguish the two effects.

In this study, we demonstrate the detection of cyclotron resonance in a graphene two-terminal device using the photo-Nernst effect under infrared light irradiation. This device was fabricated with a metal mask that covered it partially. Schematic illustrations of the fabricated device are shown in Figs. 2(a-c). Here, we briefly explain the method of device fabrication and measurement. The stack of $h$-BN/graphene/$h$-BN vdW heterostructure was fabricated on 290 nm thick $SiO_2$/doped Si substrate by using dry release transfer method [18]. The thicknesses of top and bottom $h$-BN were chosen as ~30 nm. Using e-beam (EB) lithography and EB evaporation, Pd contact with a thickness of 15 nm was fabricated. Another ~30 nm thick $h$-BN was then dry transferred on the device to cover the channel and Pd contact. With another set of EB lithography and EB evaporation, we fabricated ~100 nm thick Pd metal mask and electrodes that were extended to the contact pad. The optical micrograph of the fabricated device is show in Fig. 2(d). The measurement was performed in a variable temperature cryostat at ~3 K, which contained a superconducting magnet with the magnetic field perpendicular to the plane. Using an optical fiber and a polished metal pipe, an infrared $CO_2$ laser of wavelength $\lambda$ that ranged from 9.25 to 10.61 μm (corresponding to the photon energy range of $E_{ph}$ = 0.117–0.134 eV) was irradiated on the sample, where the laser spot size was ~33 mm$^2$ and intensity was 4.52×10$^{-2}$ Wcm$^{-2}$. We assume that the irradiated light at the sample is non-polarized as it travels through the optical fiber and the metal pipe. The photo-induced voltage was measured by the two-terminal configuration, as depicted in Fig. 2(a) using a lock-in amplifier and an optical chopper with the frequency of 108 Hz. The optical chopper modulated the incident light and provided a synchronous reference signal to the lock-in amplifier. The carrier concentration of graphene $n$ was controlled by applying the back gate voltage $V_{BG}$ to the doped-Si substrate.

Light was irradiated on only one half of the graphene channel due to the presence of the metal mask, as shown in Fig. 2(c), which illustrates the cross-sectional view of the dashed line in Fig. 2(b). Similar to Fig. 1(b), the light irradiation creates a temperature gradient around the metal mask and its direction is perpendicular to the voltage probe contacts. Fig. 2(e) shows the carrier density dependence of the photo-induced voltage $V_{ind}$ measured at a fixed magnetic field of $B = 0.82$ T, and Fig. 2(f) shows the magnetic field dependence of $V_{ind}$ measured at a fixed carrier density $n = -1 \times 10^{11}$ cm$^{-2}$. These were measured at ~3 K with $\lambda = 9.552$ μm ($E_{ph} = 0.13$ eV). The carrier density dependence of $V_{ind}$ in Fig. 2(e) demonstrates an oscillatory trend with a positive peak at charge neutrality, which is consistent with Fig. 1(d). Notably, $V_{ind}$ is almost symmetric with respect to the carrier type. This is a clear signature of the photo-Nernst effect. For the magnetic field dependence in Fig. 2(f), there is a clear asymmetry in $V_{ind}$ with respect to $B$, i.e., the $V_{ind}$ increases for positive values of $B$ and decreases for negative values of $B$. This is consistent with the observation in Fig. 1(e). Furthermore, $V_{ind}$ exhibits an oscillatory variation due to the Landau quantization of graphene. The sign of the peak and dips in $V_{ind}$ is reversed under the reversal of $B$. At $B = 0.6$, 0.75, and 0.9 T, $V_{ind}$ exhibits peaks as indicated by the solid circle, square, and triangle, respectively, whereas it exhibits dips at $B = -0.6, -0.75$, and $-0.9$ T. Overall, this response implies that $V_{ind}(-B) = -V_{ind}(+B)$. In comparison with Fig. 1, the data presented in Fig. 2(e) and 2(f) demonstrate the dominant contribution of the photo-Nernst effect in our device.

A detailed dependence of $V_{ind}$ on $B$ and $n$ at ~3 K is presented in Fig. 3(a). Here, $\lambda$ = 9.552 μm ($E_{ph}$ = 0.13 eV). This dependence reveals the Landau fan diagram of the device. In large $B$ regions, $V_{ind}$ is strongly enhanced at a particular magnetic field indicated by red arrows in the figure; This is due to the enhancement of light absorption at cyclotron resonance of graphene. Fig. 3(b) shows $V_{ind}$ vs. $n$ around the resonance magnetic field $B$ = ±1.37 T as well as at $B$ = 0. The oscillatory variation of $V_{ind}$ with respect to $n$ at $B$ = ±1.37 T is due to the carrier redistribution in Landau-quantized graphene. Dashed lines indicate the locations of the Landau level filling factors of $\nu$ = −6, −2, +2, and +6 determined from the magneto transport of two-terminal resistance of graphene (not shown). The observed oscillation period shows a good agreement with that of the Landau level filling. The peaks and dips in $V_{ind}$ at $B$ = 1.37 T are reversed in sign at $B$ = −1.37 T. Additionally, the photovoltage signal is negligible at zero magnetic field. Consistent with Figs. 1(e) and 1(f), the signal at resonance satisfies the condition of the photo-Nernst effect. Furthermore, we measured cyclotron resonance in the large magnetic field region as shown in Fig. 3(c). The left panel of Fig. 3(c) showing dependence of $V_{ind}$ on $B$ in the large magnetic field region. Here, $\lambda$ = 9.552 μm ($E_{ph}$ = 0.13 eV). The cross-section of the data along $B$ at a fixed carrier density $n$ = 1.58 ×10$^{11}$ cm$^{-2}$ (this value is indicated by the black dashed line in the left panel) is presented in the right panel. Data exhibits peak around ~±10 T and this high field resonance also exhibits sign reversal with respect to the magnetic field direction.

We now discuss the dependence of the observed resonance peaks on photon energy. When incident energy increases, the resonance field moves toward larger magnetic field, as shown in Fig. 4(a). To obtain further insights, we compare this trend with the magnetic field dependence of the Landau-level energy in monolayer graphene, as shown in Fig. 4(b).

In the presence of the magnetic field, the electron energy of graphene changes according to $E_N = v_F\sqrt{2e\hbar NB}$, where $v_F$ denotes the Fermi velocity of graphene, $e$ the elementary charge, $\hbar$ the reduced Planck constant, $B$ the magnetic field perpendicular to graphene, and $N$ the Landau level index. Energy levels from $N = 0$ to $N = \pm 2$ are plotted in this figure. When the energy of incident light satisfies the selection rule of graphene ($\Delta|N| = \pm 1$), light absorption occurs, which corresponds to cyclotron resonance. The cyclotron resonance field varies with the laser energy. The observed magnetic field resonance and laser energy corresponds to the transition from $N = -1$ to 2 and $-2$ to 1; these transitions are depicted by red arrows in the figure. Figure 4(c) shows this transition energy, indicated by a red dashed line, along with the experimentally observed resonance magnetic field vs. laser energy, depicted as red solid squares. A good agreement is observed between the theoretical and experimental results for $v_F \sim 1.1 \times 10^6$ m/s. This $v_F$ value is consistent with that typically observed for $h$-BN encapsulated graphene [19]. These results confirm that we have detected the cyclotron resonance in graphene by using the photo-Nernst effect. We also note that another resonance around ~10 T in Fig.3(c) corresponds to the transition from $N = 0$ to 1 and $-1$ to 0 in Fig. 4(b). The variation of this resonance field is also plotted in Fig. 4(c) indicated by black solid circles, which can be explained by the cyclotron resonance using the same $v_F$ value. As the cyclotron resonance is essentially the enhancement of the optical absorption of graphene, our results unambiguously demonstrate that the obtained photovoltage originated from the optical absorption of the graphene. This rules out the effect of infrared absorption of the substrate that induces the heating of graphene [20].

In conclusion, we have demonstrated the detection of cyclotron resonance using the photo-Nernst effect under infrared irradiation. The observed asymmetry of the photovoltage with respect to the sign of the magnetic field can be fully explained with the photo-Nernst effect without considering other effects. The sensitivity of the photo-Nernst effect at cyclotron resonance was obtained as 81 V/W by restricting the laser at the uncovered graphene region. This sensitivity is similar to that obtained for the photo-Seebeck effect in graphene [14,21] and establishes the high sensitivity of the photo-Nernst effect. The photo-Nernst effect provides another mechanism for sensitive far-infrared light detection.

**Acknowledgements**

This work was supported by CREST, Japan Science and Technology Agency (JST) (grant number JPMJCR15F3), and JSPS KAKENHI (grant numbers JP19H02542 and JP19H01820).

**Figure captions**

Figure 1

Schematic illustrations of (a-e) photo-Nernst effect and (f-j) photo-Seebeck effect. Schematic structure of a device exhibiting (a) photo-Nernst effect and (f) photo-Seebeck effect. (b,g) Temperature gradient generated under light irradiation. Graphene's carrier density $n$ dependence of (c,d) photo-Nernst and (h,i) photo-Seebeck signal. Illustrations are presented for both (c,h) zero magnetic field and (d,i) high magnetic fields. (e,j) Magnetic field dependence of (e) photo-Nernst and (j) photo-Seebeck signal.

Figure 2

(a,b) Schematic illustrations of (a) side-view and (b) top-view of $h$-BN/graphene/$h$-BN device with a metal mask. Metal contacts and electrical contacts for detecting photovoltage and applying back gate voltage $V_{BG}$ are shown here. Magnetic field is applied perpendicular to the plane. (c) Cross-sectional view of the device. (d) Optical micrograph of the device. (e) Dependence of the photo-induced voltage $V_{ind}$ on carrier density measured at a fixed magnetic field $B = 0.82$ T. (f) Dependence of the photo-induced voltage $V_{ind}$ on magnetic field measured at fixed carrier density $n = -1 \times 10^{11}$ cm$^{-2}$. These measurements were performed at 3.0 K, where the laser with wavelength of $\lambda = 9.552$ μm is irradiated. The locations of peaks and dips in $V_{ind}$ are marked by circles, squares, and triangles. A constant background signal of ~100 nV was present during the measurement and this signal is subtracted from the data.

Figure 3

(a) Dependence of $V_{ind}$ on magnetic field $B$ and carrier density $n$ at 3.0 K. Here, $\lambda = 9.552$ μm. (b) Dependence of $V_{ind}$ on carrier density $n$ for $B = +1.37$, 0, and $-1.37$ T. The individual traces are offset for clarity and the offset levels for each of them are indicated by horizontal dotted lines. A vertical dashed line depicts the position of Landau level filling factors $\nu = -6, -2, +2$ and $+6$. (c) Left-panel: High magnetic field behavior of $V_{ind}$ as a function of $B$ and $n$ at 3.0 K with $\lambda = 9.552$ μm. Right-panel: cross-section of left panel along $B$ at a fixed $n = 1.58 \times 10^{11}$ cm$^{-2}$. A constant background signal of ~100 nV is subtracted from the data in panels (a-c).

Figure 4

(a) Dependence of $V_{ind}$ on magnetic field $B$ for different energies of the incident light. (b) Energy diagrams of the electrons in graphene in the presence of a perpendicular magnetic field. Coincident conditions of inter-Landau-level transition and optical excitation ($\lambda = 9.552$ μm) are depicted by red and black arrows for $N = -1\ (-2) \rightarrow N = +2\ (+1)$ and $N = 0\ (-1) \rightarrow N = +1\ (0)$ transitions, respectively. (c) Energy of the incident light as a function of resonance magnetic field. Solid squares and circles indicate experimental data for resonances in low and high magnetic field regions, respectively. Dashed line indicates the calculated transition energy between different Landau levels using Fermi velocity of monolayer graphene of $v_F \sim 1.1 \times 10^6$ m/s.

Figure 1

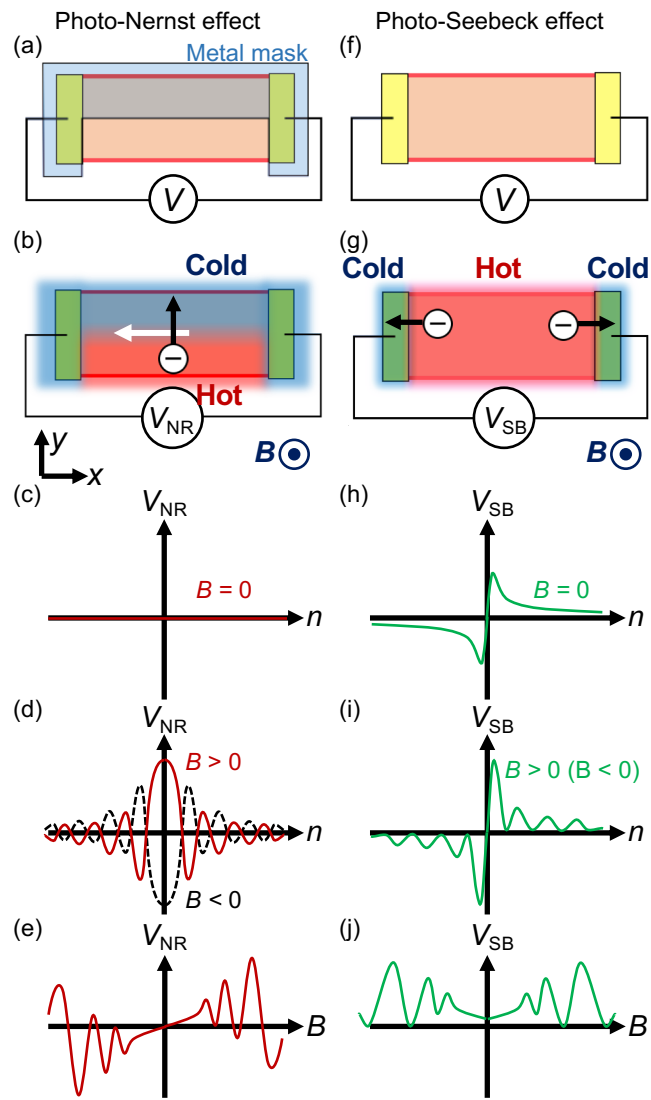

Figure 2

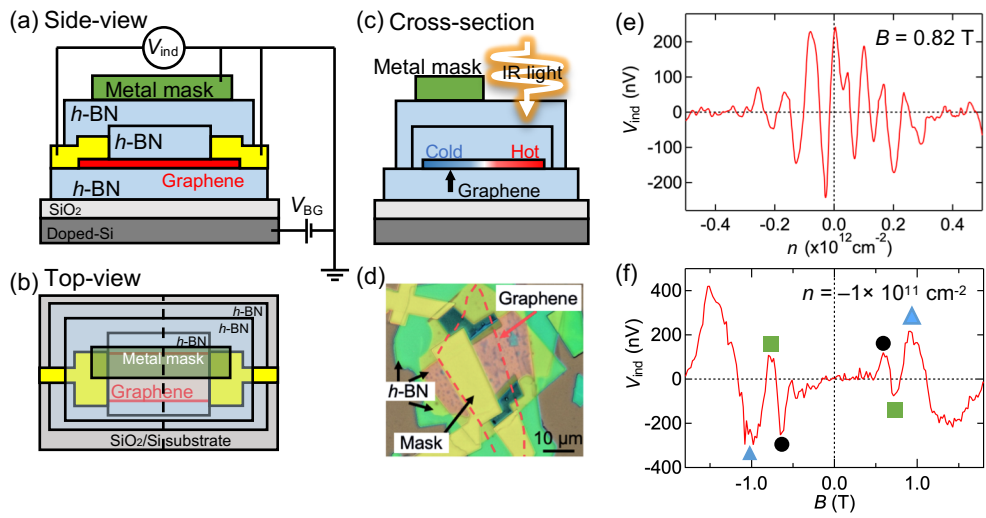

Figure 3

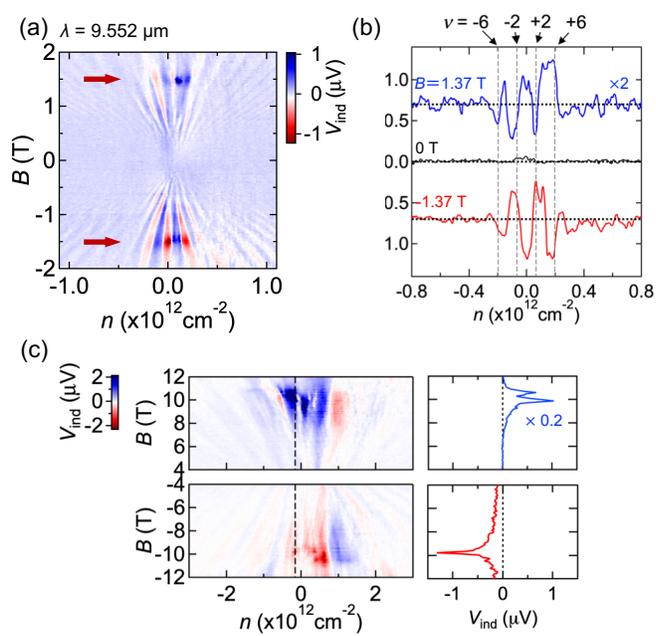

Figure 4

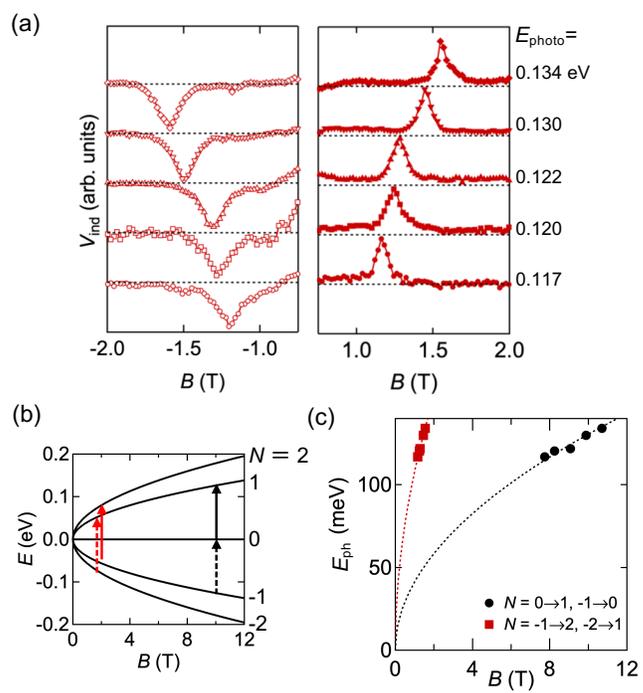